\documentclass[11pt]{article}
\usepackage[dvips]{color}
\usepackage{epsfig}
\usepackage{amsmath}
\usepackage{graphicx}
\date{}
\begin{document}
\title{{\bf Classical and quantum dynamics of a perfect
fluid scalar-energy dependent metric cosmology}}

\author{M. Khodadi$^1$\thanks{%
m.khodadi@stu.umz.ac.ir},\,\, K. Nozari$^1$\thanks{%
knozari@umz.ac.ir},\,\, and B. Vakili$^2$\thanks{%
b.vakili@iauctb.ac.ir (Corresponding author)}
\\\\
$^1${\small {\it Department of Physics, Faculty of Basic Sciences,
University of Mazandaran,}}\\ {\small {\it P.O. Box
47416-95447, Babolsar, Iran}} \\
$^2${\small {\it Department of Physics, Central Tehran Branch,
Islamic Azad University, Tehran, Iran}}}

\maketitle

\begin{abstract}
Inspired from the idea of minimally coupling of a real scalar field
to geometry, we investigate the classical and quantum models of a
flat energy-dependent FRW cosmology coupled to a perfect fluid in
the framework of the scalar-rainbow metric gravity. We use the
standard Schutz' representation for the perfect fluid and show
that under a
particular energy-dependent gauge fixing, it may lead to the
identification of a time parameter for the corresponding dynamical
system. It is shown that, under some circumstances on the
minisuperspace prob energy, the classical evolution of the of the
universe represents a late time expansion coming from a bounce
instead of the big-bang singularity. Then we go forward by showing
that this formalism gives rise to a Schr\"{o}dinger-Wheeler-DeWitt
(SWD) equation for the quantum-mechanical description of the model
under consideration, the eigenfunctions of which can be used to
construct the wave function of the universe. We use the resulting
wave function in order to investigate the possibility of the
avoidance of classical singularities due to quantum effects by means
of the many-worlds and Bohmian interpretation of quantum cosmology.\vspace{5mm}\noindent \vspace{0.8mm}\newline

\end{abstract}
\section{Introduction}
One of the most important questions in cosmology is that of the
initial conditions from which our present universe has began its
evolution. As is well known, any answer to this question according
to the standard model of cosmology based on the classical general
relativity (GR), suffers from the existences of an initial
singularity, the so-called big-bang singularity. Therefore, removing
this initial singularity from the cosmological solutions of the
Einstein's equations is one of the most important challenges of GR
and any hope to dealing with this issue would be in development of a
quantum theory of gravity and consequently quantum cosmology (QC).
In QC a common process to remove the big-bang is to replace it with
a bounce in the sense that instead of initial singular "bang", the
universe contracts until a minimal size (usually of the order of the
Planck size $l_{Pl}$), bounces from this minimal size and then
re-expands in such a way its late time behavior is given by
classical cosmology. However, if one accepts that such a minimal
size might be a fundamental quantity, with an eye to special
relativity, this question may be arise: how is it possible to have a
basic minimal length in nature which all inertia observers agree on
its value. One may seek the answer to this question in one of the
main motivations for the formation of an effective approach to QG
knows as ``Doubly Special Relativity'' (DSR) \cite{G1, G2}. In DSR,
in addition of the speed of light in standard SR, the non-linear
representation of Lorentz transformations in momentum space gives an
other global constant as Planck energy  \cite{G3, G4, J, Mag1,
Mag2}. Concretely speaking, by fixing the natural cutoffs as planck
length (or energy) DSR plays the role of a flat space-time limit of
QG \cite{Jon}. As a result, in the presence of the non-linear
Lorentz symmetry, standard dispersion relations may be modified up
to the leading order of Planck length \cite{M, Ja, Am, Co, Ha}. A
natural extension of DSR is when gravity (the curvature of
space-time) comes into the play by which we are led to `` Doubly
General Relativity'' (DGR) or ``gravity's rainbow'' which is based
on two principles: the correspondence principle and the deformed
equivalence principle \cite{Smolin}. The most important feature of
this proposal which lead to breakdown of standard GR is that from
the perspective of the particle(s) which is (are) exploring the
space-time, the single classical geometry does not exist. This is a
direct consequence of the disruption of the system under the
measurement (here the geometry of space-time) by means of the
measuring tools (here prob particle(s)). More technically, in the
framework of the gravity's rainbow, the geometry of space-time is
described by a rainbow metric which itself is composed of one
parameter family of the metrics which are functions of the energy of
the prob particle(s). In the recent years, this semi-classical
formalism of QG has attracted much attentions as far as various
aspects of its cosmological implication has been studied, see
\cite{RG1}-\cite{Y2}.

In this paper we are going to extend a previously studied work on
the perfect fluid scalar-metric cosmology \cite{Babak}, and
investigate the classical and quantum dynamical evolution of a flat
energy-dependent FRW cosmology with a perfect fluid matter source
and a non-linear self-coupling scalar field minimally coupled to
gravity's rainbow. To express clear, in this work we would like to
derive cosmological results in the presence of a naturally cutoff as
Planck energy for FRW space-time metric. For the matter source of
gravity, we consider a perfect fluid in Schutz' formalism \cite{Sc1,
Sc2}. The advantage of using this formalism is that after a
canonical transformation the conjugate momentum associated to one of
the variables of the fluid appears linearly in the Hamiltonian and
so, in a natural way, it can offer a time parameter in terms of
dynamical variables of the perfect fluid.  Therefore, canonical
quantization results in a Schr\"{o}dinger-Wheeler-DeWitt (SWD)
equation, in which this matter variable plays the role of time
\cite{Bar}-\cite{ani}. The paper has the following structure. In
section 2, we present a brief review about the construction process
of the rainbow geometry so that in the end we introduce the
energy-dependent isotropic and homogeneous FRW space-time metric. In
section 3, in terms of the Schutz' time parameter, we shall obtain
the classical dynamical behavior of the rainbow cosmic scale factor
and the scalar field. We deal with the quantization of the model in
section 4, and by computing the expectation values of the scale
factor and the scalar field and also their Bohmian (ontological)
counterparts in section 5, we show that the evolution of the
universe according to the quantum picture is free of classical
singularities. Section 6 is devoted to the summary and conclusions.

\section{Rainbow geometry construction}
As mentioned above, the main idea of DGR is based on the
correspondence principle and the deformed equivalence principle.
Therefore to construct the geometric structure corresponding to the
rainbow scenario, checking these two principles is absolutely
necessary. According to the correspondence principle, in the limit
of low energy scales compared with $E_{Pl}$, the standard GR should
be recovered. This means that
\begin{equation}\label{e2-1}
\lim_{\frac{E}{E_{Pl}}\rightarrow0}g_{\mu\nu}(E)=g_{\mu\nu}~.
\end{equation}
It should be noted that the dependence of the space-time metric on
the energy may be interpreted as the energy scale in which the
geometry of space-time is explored by means of test particle(s).
Deformed equivalence principle, on the other hand, can be explained
as: consider free falling observers in a region of space-time with a
radius of curvature $R\gg l_{Pl}$, which measure physical objects
such as particles and fields with energy $E$, the laws of physics
are the same, provided that
\begin{equation}\label{e2-2}
R^{-1}\ll E\ll E_{Pl}~.
\end{equation}
The above condition guarantees that based on DGR scenario, in the
absence of the gravity (locally) DSR is recovered but not the
standard SR. In summary, the deformed equivalence principle implies
that in a region of the size $l_{Pl}$ around a given point of
space-time, it is always possible to find a coordinate system in
which the effects of gravity are negligible. Such a coordinate
system is called ``local deformed Lorentz frame''. By considering
the above mentioned free falling observers as inertial observers in
a rainbow flat space-time, we offer a family of local energy
dependent orthonormal frame fields as
\begin{equation}\label{e2-3}
e_{0}=f_{1}^{-1}(\frac{E}{E_{Pl}})\tilde{e}_{0},~~~~~~ e_{i}=
f_{2}^{-1}(\frac{E}{E_{Pl}})\tilde{e}_{i},
~~~(i=1,2,3)~,
\end{equation}in which $\tilde{e}_{\mu}$ are the standard GR
counterparts of $e_{\mu}$. Therefore, one gets the following flat
metric for all inertial observers with various energies $E$
\begin{equation}\label{e2-4}
g(E)=\eta^{\mu \nu}e_{\mu}\otimes e_{\nu}~.
\end{equation}
Here, $f_1$ and $f_2$ are two unknown energy dependent functions
known as ``rainbow functions'' which according to the correspondence
principle satisfy the limiting relations:
$\lim_{\frac{E}{E_{Pl}}\rightarrow0}
f_{1,2}(\frac{E}{E_{Pl}})\rightarrow1$. This means that in the low
energy limit we have $e_{\mu}\rightarrow\tilde{e}_{\mu}$ which it
seems to be quite reasonable. It is important to note that due to
freedom in choosing the functions $f_{1,2}(\frac{E}{E_{Pl}})$, it is
always possible to have functions with a strong damping as gaussian
for example. This means that although we have introduced the
interval (\ref{e2-2}) as an restriction on the energy of the probing
particles, there is no reason to restrict the investigation with a
cut-off at the Planck scale. Indeed, we may relax such a restriction
and look for the rainbow particles with trans-Planckian energies
\cite{Ga}. Particulary, a gaussian profile for the rainbow functions
can have physical implications, see for instance \cite{Ga}, in which
by using of a variational viewpoint based on gaussian trial
functionals, it is shown that some typical divergences can be
controlled.

Metric (\ref{e2-4}) is known as flat
DSR space-time or deformed Minkowski space-time and in terms of its
components may be written as
\begin{equation}\label{e2-5}
dS^2 =\frac{(dx^{0})^{2}}{f_{1}^{2}(\frac{E}{E_{Pl}})}-\frac{
(dx^{i})^{2}}{f_{2}^{2}(\frac{E}{E_{Pl}})},
\end{equation}
which $x^{0}$ and $x^{i}$ denote the time and spatial coordinates
respectively.

Now, let us extend this issue to the cases in which the space-time
metric has a dynamical behavior. By such an extension to a
homogeneous and isotropic cosmological model, we may introduce the
following energy-dependent (rainbow) FRW metric
\begin{equation}\label{e2-a}
dS^2 = \frac{N^2(t)}{f_1^2(\varepsilon)} dt^2 -
\frac{a^2(t)}{f_2^2(\varepsilon)} \left[\frac{dr^2}{1-kr^2} + r^2
d\vartheta ^2 + r^2 \sin^2 \vartheta d\varphi^2 \right],
\end{equation}
where, as usual, $N(t)$ is the lapse function, $a(t)$ the scale
factor and $k=-1,0,1$ corresponds to the open, flat and closed
universe respectively. Also, the the dimensionless parameter
$0<\varepsilon <1$ is defined as $\varepsilon=\frac{E}{E_{Pl}}$. \\
In what follows, we will study the classical and quantum evolution
of a FRW universe in its radiation dominate period in the context
of rainbow gravity. To do this, due to the statistical nature of the
very early universe, we take an ensemble of massless particles (like
photons or gravitons) as prob particles of the space-time. In fact,
we want to consider the average effect of these particles (which are
moving on the space-time geometry) during a given measuring process.
We will see that the average energy of the prob photons may vary or
may be constant with the time evolution of universe. In the following
sections, the implications of the both possibilities will be considered.

\section{The classical cosmological model}
Let us start with a FRW cosmological model whose metric is coupled
with a scalar field. Also, the matter role of our model is played
with a perfect fluid. The action of such a structure may be written
as (we work in units where $c=\hbar=16\pi G=1$)
\begin{equation}\label{e3-1}
{\cal S}=\int_M d^4x \sqrt{-g}\left[R-F(\phi)g^{\mu \nu}\phi_{,\mu}
\phi_{,\nu}\right]+2\int_{\partial M}d^3x \sqrt{h}h_{ab}K^{ab}+
\int_M d^4 x \sqrt{-g}p,
\end{equation}
where $R$ is the Ricci scalar and $F(\phi)$ is an arbitrary function
of the scalar field $\phi$. Also, $K^{ab}$ and $h_{ab}$ are the
extrinsic curvature and induced metric on the three-dimensional
spatial hypersurfaces, respectively. The boundary term in
(\ref{e3-1}) will be canceled by the variation of $\int_{M}
d^4x\sqrt{-g}R$ \footnote{With a straightforward calculation one can
show that the variation of this term produces the boundary integral
$-\int_{\partial M}\sqrt{h}\epsilon h^{\alpha \beta}\delta g_{\alpha
\beta ,\mu}n^{\mu}d^3y$, where $n_{\mu}$ is the unit normal to
$\partial M$ and $\epsilon=n_{\mu}n^{\mu}=\pm1$, which is exactly
equal to the variation of the second term in (\ref{e3-1}) but with
different sign.}. This is the reason for including a boundary term
in the action of a gravitational theory. That such a boundary term
is needed is due to the fact that $R$, the gravitational Lagrangian
density contains second derivatives of the metric tensor, a
nontypical feature of field theories. The third term of (\ref{e3-1})
represents the matter contribution to the total action where $p$ is
the pressure of the fluid which together with its energy density
$\rho$ obey the following equation of state
\begin{equation}\label{e3-2}
p=\omega \rho.
\end{equation}
A more familiar expression for the action of a perfect fluid is the
Hawking-Ellis action $S_{P.F.}=-\int d^4x \sqrt{-g}
\zeta(1+\upsilon)$, in which $\zeta$ and $\upsilon$ are the fluid's
density and elastic potential (or internal energy) respectively
\cite{Ellis}. By definition of the fluid's energy density and
pressure as $\rho=\zeta(1+\upsilon)$ and
$p=\zeta^2\frac{d\upsilon}{d\zeta}$, it is shown in \cite{Ellis}
that the variation of this action with respect to the metric yields
the standard expression $T^{\mu \nu}=(\rho+p)U^{\mu}U^{\nu}+pg^{\mu
\nu}$, for the energy-momentum tensor of a perfect fluid. However,
according to what is shown in \cite{Sc1, Sc2}, the same result may
be obtained if one varies the action $\int d^4x \sqrt{-g}p$ with
respect to the metric. In this respect, the matter part of the
action as is written in terms of pressure in (\ref{e3-1}) is
equivalent to the usual Hawking-Ellis formalism for the perfect
fluid.

In this work we consider the equation of state parameter $\omega$ as
a constant whose values vary between $-1\leq \omega \leq 1$.
According to the Schutz's formalism the fluid's four-velocity can be
expressed in terms of some thermodynamical potentials as \cite{Sc1,
Sc2}
\begin{equation}\label{e3-3}
U_{\nu} = \frac{1}{\mu} (\epsilon_{,\nu} + \theta S_{,\nu}),
\end{equation}
where $\mu$ and $S$ indicate the specific enthalpy and the specific
entropy, respectively. Also, the variables $\epsilon$ and $\theta$
have no clear physical interpretation in this formalism. Note that
the fluid's four velocity should satisfy the normalization condition
$U_{\nu}U^{\nu} =-1$. Now, considering the above action as a
dynamical system in which the scale factor $a$, scalar field $\phi$
and fluid's potentials are considered as independent dynamical
variables, we can rewrite the gravitational part of the action
(\ref{e3-1}) as

\begin{eqnarray}\label{e3-6}
{\cal S}_{g}&=&\int dt {\cal L}_{g}= \int
dt\frac{Na^{3}}{f_{1}f_{2}^{3}}dt
\left\{\bigg(R-\frac{f_{1}^{2}}{N^{2}}\dot{\phi}^{2}F(\phi)\bigg)\right.
\nonumber\\ &
&\left.-\lambda\left[R-\frac{6}{N^2}\left(\frac{\ddot{a}}{a}+
\frac{f_{1}^{2}\dot{a}^2}{a^2}+\frac{kf_{2}^{2}}{a^2}-\frac{\dot{N}\dot{a}}
{Na}\right)\right]\right\}~,
\end{eqnarray}in which we have used the definition of the Ricci
scalar in terms of the metric functions and their derivatives as a
constraint. In this procedure the Lagrange multiplier $\lambda$ can
be obtained by variation with respect to $R$, with the result
$\lambda =\frac{Na^{3}}{f_{1}f_{2}^{3}}$.  Thus, one obtains the
following point-like Lagrangian for the gravitational part of the
model
\begin{eqnarray}\label{e3-7a}
{\cal L}_{g}=\frac{f_{1}}{f_{2}^{3}}\frac{6a\dot{a}^{2}}{N}+
\frac{6kNa}{f_{1}f_{2}}-\frac{f_{1}}{Nf_{2}^{3}}F(\phi)\dot{\phi}^{2}
a^{3}~.
\end{eqnarray}
Now, let us return to the matter part of the action (\ref{e3-1})
from which the Lagrangian density of the fluid takes the form
\begin{eqnarray}\label{e3-7}
{\cal L_{}}_{m}=\frac{Na^3}{f_{1}f_{2}^{3}}p~.
\end{eqnarray}
By using of the first law of thermodynamics and following the
standard Schutz's description of the perfect fluid it can be shown that
the fluid's equation of state is of the form (see \cite{Babak}
for details)
\begin{equation}\label{e3-11}
p=\frac{\omega}{(\omega+1)^{\frac{\omega+1}{\omega}}}\mu^{\frac
{\omega+1}{\omega}}e^{-\frac{S}{\omega}}~.
\end{equation}
On the other hand if we use the expression
$U_{\nu}=(\frac{N}{f_{1}(\varepsilon)},0,0,0)$ as the fluid's
four-velocity in the normalization condition $U_{\nu}U^{\nu} =-1$,
we get
\begin{equation}\label{e3-12}
\mu=\frac{(\dot{\epsilon}+\theta \dot{S})}{N}f_{1}(\varepsilon)~.
\end{equation}
Finally, using the above constraints and thermodynamical
considerations for the fluid we find the Lagrangian density of the
matter as
\begin{equation}\label{e3-13}
{\cal
L}_{m}=\frac{f_{1}^{\frac{1}{\omega}}}{f_{2}^{3}}N^{-\frac{1}{\omega}}
a^3\frac{\omega}{(\omega+1)^{1+\frac{1}{\omega}}}\left(\dot{\epsilon}+
\theta\dot{S}\right)^{1+1/\omega}e^{-\frac{S}{\omega}}~.
\end{equation}
We are now in a situation to construct the Hamiltonian for the
model. In terms of the conjugate momenta the Hamiltonian is given by
\begin{equation}\label{e3-14}
H=\dot{a}P_a+\dot{\phi}P_{\phi}+\dot{\epsilon}P_{\epsilon}+\dot{S}P_S-{\cal
L}~,
\end{equation}
where $H= H_{g}+H_{m}$ and ${\cal L}={\cal L}_{g}+{\cal L}_{m}$. The
momenta conjugate to each of the above variables can be obtained
from the definition $P_{q}=\frac{\partial {\cal L}}{\partial
\dot{q}}$ with the result
\begin{eqnarray}\label{e3-15}
\left\{
\begin{array}{ll}
P_{a}=\frac{12f_{1}(\varepsilon)}{f_{2}^{3}(\varepsilon)}\frac{\dot{a}}{N}~,\\\\
P_{\phi}=-\frac{12f_{1}(\varepsilon)}{f_{2}^{3}(\varepsilon)}F(\phi)\dot{\phi}a^{3}~,\\\\
P_{\epsilon}=\frac{f_{1}^{\frac{1}{\omega}}(\varepsilon)}{f_{2}^{3}(\varepsilon)}N^{-\frac{1}
{\omega}}a^3\frac{\omega(1+\frac{1}{\omega})}{(\omega+1)^{1+\frac{1}{\omega}}}\left
(\dot{\epsilon}+\theta\dot{S}\right)^{\frac{1}{\omega}}e^{-\frac{S}{\omega}}~,\\\\
P_{S}=\frac{f_{1}^{\frac{1}{\omega}}(\varepsilon)}{f_{2}^{3}(\varepsilon)}N^{-\frac{1}{\omega}}a^3
\frac{\omega(1+\frac{1}{\omega})\theta}{(\omega+1)^{1+\frac{1}{\omega}}}\left
(\dot{\epsilon}+\theta\dot{S}\right)^{\frac{1}{\omega}}e^{-\frac{S}{\omega}}~.
\end{array}
\right.
\end{eqnarray}With the help of these relations and also applying the canonical
transformation \cite{V.G}
\begin{equation}\label{e3-17}
T=-P_Se^{-S}P_{\epsilon}^{-(\omega+1)}~,~~~~~~~
P_T=P_{\epsilon}^{\omega+1}e^S~,
\end{equation}
expression (\ref{e3-14}) leads to \footnote{It should be noted that
while the fluid's enthalpy depends on the rainbow function $f_1$
through equation (\ref{e3-12}), its entropy seems to be free of such
dependence. However, due to the Bekenstein-Hawking area law which is
also applicable to the cosmological horizons, the entropy may also
depend on the metric (and so to the rainbow) functions. In this
situation, we may absorb all of such dependencies in variable $S$.
This means that in the above discussions, by $S$ we mean a metric
dependence function. Finally, by applying of the canonical
transformation (\ref{e3-17}), all information will be delivered to
the variable $T$.}
\begin{equation}\label{e3-16}
H=N{\cal H}= \frac{N}{f_1(\varepsilon)}
\left[-\frac{f_2^3(\varepsilon)}{24a} p_a^2-
\frac{6ka}{f_2(\varepsilon)}
+\frac{f_2^3(\varepsilon)}{4F(\phi)a^{3}}P_{\phi}^{2}+
\frac{f_2^{3\omega}(\varepsilon)}{a^{3\omega}}P_{T} \right] ~.
\end{equation}
The momentum $P_T$, which in comparison with the other momenta in
the Hamiltonian function appears linearly, is the only remaining
canonical variable associated with the matter.

The classical equations of motion are governed in the Hamiltonian
formalism as
\begin{eqnarray}\label{e3-18}
\left\{
\begin{array}{ll}
\dot{a}=\{a,H\}=-\frac{N}{12}\frac{P_a}{a}~\frac{f_2^3(\varepsilon)}{f_{1}(\varepsilon)}~,\\\\
\dot{P_a}=\{P_a,H\}=N\left[-\frac{1}{24}\frac{P_a^2}{a^2f_{1}(\varepsilon)}+\frac{3f_2^3(\varepsilon)}
{4F(\phi)a^4f_1(\varepsilon)}P_{\phi}^2+\frac{6k}{f_{1}(\varepsilon)f_2(\varepsilon)}
+3\omega \frac{f_2^{3\omega}(\varepsilon)}{f_{1}(\varepsilon)}a^{-3\omega-1}P_T\right]~,\\\\
\dot{\phi}=\{\phi,H\}=\frac{Nf_2^3(\varepsilon)}{2F(\phi)a^3f_{1}(\varepsilon)}P_{\phi}~,\\\\
\dot{P_{\phi}}=\{P_{\phi},H\}=N\frac{f_2^3(\varepsilon)P_{\phi}^2}{4a^3f_1(\varepsilon)}\frac
{F'(\phi)}{F(\phi)^2}~,\\\\
\dot{T}=\{T,H\}=\frac{N}{a^{3\omega}}~\frac{f_2^{3\omega}(\varepsilon)}{f_{1}(\varepsilon)}~,\\\\
\dot{P_T}=\{P_T,H\}=0~.
\end{array}
\right.
\end{eqnarray}
We also have the Hamiltonian constraint ${\cal H}=0$. Up to now, the
cosmological setting, in view of the concerning issue of time, has
been of course under-determined. Before dealing with the solutions
of these equations we must decide on a choice of time parameter
which at the classical level may be resolved by using the gauge
freedom via fixing the gauge. A glance at the above equations shows
that by choosing the gauge $
N=\frac{f_{1}(\varepsilon)}{f_2^{3\omega}(\varepsilon)}a^{3\omega}$,
we have $T=t$, which means that variable $T$ may play the role of
time. Now, let us rewrite the equations of motion in the gauge $
N=\frac{f_{1}(\varepsilon)}{f_2^{3\omega}(\varepsilon)}a^{3\omega}$
as
\begin{eqnarray}\label{e3-19}
\left\{
\begin{array}{ll}
\dot{a}=-\frac{a^{3\omega-1}}{12}f_2^{3-3\omega}(\varepsilon)P_a~,\\\\
\dot{P_a}=-\frac{a^{3\omega-2}}{24}f_{2}^{-3\omega}(\varepsilon)P_a^2+
\frac{3f_2^{3-3\omega}(\varepsilon)a^{3-3\omega}}{4F(\phi)a^4f_1(\varepsilon)}
P_{\phi}^2+6ka^{3\omega}f_2^{-3\omega-1}(\varepsilon)+3\omega a^{-1}P_{0}~,\\\\
\dot{\phi}=\frac{a^{3\omega-3}f_{2}^{-3+3\omega}(\varepsilon)}{2F(\phi)}P_{\phi}~,\\\\
\dot{P_{\phi}}=\frac{F'(\phi)}{4F}a^{3\omega-3}f_{2}^{3-3\omega}(\varepsilon)P(\phi)^2~.
\end{array}
\right.
\end{eqnarray}
Here, we take $P_{T}=P_{0}$=const. By eliminating $P_{\phi}$ from
the two last equations of the above system we obtain
\begin{equation}\label{e3-20}
\frac{\ddot{\phi}}{\dot{\phi}}+\frac{1}{2}\frac{F'(\phi)}{F(\phi)}\dot{\phi}
+3(1-\omega)\left(\frac{\dot{f}_{2}(\varepsilon)}{f_{2}(\varepsilon)}+\frac
{\dot{a}}{a}\right)=0~,
\end{equation}
which can easily be integrated to yield
\begin{equation}\label{e3-21}
\dot{\phi}^2 F(\phi)=C a^{6\omega-6}f_{2}^{6\omega-6}(\varepsilon)~,
\end{equation}
where $C$ is an integration constant. Also, if we remove the momenta
from the system (\ref{e3-19}) we arrive at the differential form of
the Hamiltonian constraint as
\begin{equation}\label{e3-22}
-6a^{1-3\omega}f_{2}^{3\omega-3}(\varepsilon)\dot{a}^2+\dot{\phi}^2
F(\phi)f_{2}^{3\omega-3}(\varepsilon)a^{3-3\omega}-6kf_{2}^{-3\omega-1}
(\varepsilon)a^{1+3\omega}+P_0=0~.
\end{equation}
With the help of (\ref{e3-21}) this equation, for the flat case
$k=0$, can be put into the form
\begin{equation}\label{e3-23}
\frac{da}{dt}=\sqrt{\frac{C}{6}f_{2}^{6\omega-6}(\varepsilon) a^{6\omega-4}
+\frac{P_0}{6}f_{2}^{3-3\omega}(\varepsilon)a^{3\omega-1}}~.
\end{equation}
Before integration of this equation we have to choose the functional
form of the rainbow function $f_{2}(\varepsilon) $ in terms of scale
factor $a$. However, rainbow functions are usually a function of the
energy of the prob particles moving on the geometry of space-time.
In \cite{Adel}, by considering the early universe as a
thermodynamical system filled with prob photons at thermal
equilibrium, the relation between the average energy $\bar{E}$ of
the ensemble of prob photons and energy density $\rho$ of the
universe is obtained as $\bar{E}=\frac{4{\cal
C}}{3}\rho^{\frac{1}{4}}$ (${\cal C}$ is a constant). Therefore,
with the rainbow function of the form
$f_{2}(\varepsilon)=(1-\frac{E}{E_{Pl}})^{-1}$ offered in
\cite{Mag1}, we find
\begin{equation}\label{e3-24}
f(\rho)=\big(1-\frac{4{\cal C}}{3E_{Pl}}
\rho^{\frac{1}{4}}\big)^{-1}~.
\end{equation}
Finally, by replacing the the relation $\rho=\rho_{0}
a^{-3(1+\omega)}$ equation (\ref{e3-23}) takes the
form
\begin{equation}\label{e3-26}
\frac{da}{dt}=\sqrt{\frac{C}{6} a^{6\omega-4}+\frac{P_0}{6}a^{3\omega-1}
+ C\eta(1-\omega)a^{\frac{21}{4}\omega-\frac{19}{4}}+\frac{P_0\eta}{2}
(\omega-1)a^{\frac{9}{4}\omega-\frac{7}{4}}}~.
\end{equation}
Here, $\eta=\frac{4{\cal C}\rho_{0}^{\frac{1}{4}}}{3E_{Pl}}$ is a
small constant. It is seen that this equation can not be solved
analytically. In figure 1 (left), employing numerical methods, we
have shown the approximate behavior of the scale factor $a(t)$ for
typical values of the parameters. We see that the evolution of the
universe begins with a big-bang singularity at $t=0$ and follows an
expansion phase at late time of cosmic evolution. To understand the
relation between the big-bang singularity of the scale factor
$a\rightarrow 0$, and possible singularities of the scalar field
like the blow up singularity $\phi\rightarrow \pm \infty$, let us
find the classical trajectory in configuration space $(a,\phi)$,
where the time parameter $t$ is eliminated.
\begin{figure}
\begin{tabular}{c}\hspace{-1cm}\epsfig{figure=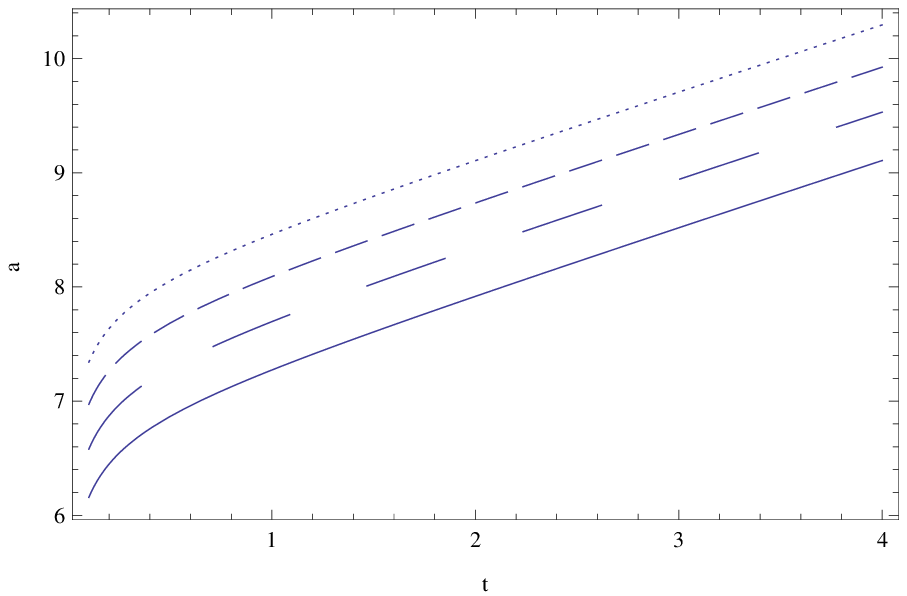,width=7cm}
\hspace{1cm} \epsfig{figure=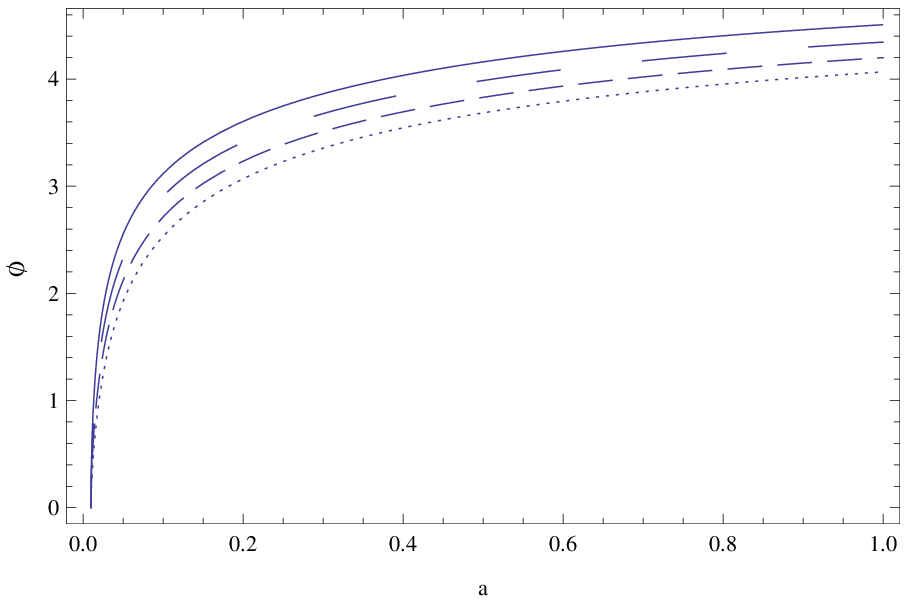,width=7cm}
\end{tabular}
\caption{\footnotesize  Left: The figure shows the time-behavior of
the classical rainbow scale factor extracted from (\ref{e3-26}). The
figure is plotted for the numerical values $\eta=0.001$,
$\eta=0.0012$, $\eta=0.0014$ and $\eta=0.0016$ from down to up.
Right: The classical trajectory in the plane $\phi-a$ for
$\eta=0.001$, $\eta=0.002$, $\eta=0.003$ and $\eta=0.004$ from up to
down. We set the numerical values $P_0= 1$, $C=2$, with the EOS
parameters $\omega=\frac{1}{3}$ for the period of the radiation
dominate.} \label{fig1}
\end{figure}
From (\ref{e3-21}), (\ref{e3-22}) one has
\begin{equation}\label{e3-27}
\frac{\sqrt{F(\phi)}d\phi}{da}=
\frac{\sqrt{C}a^{3\omega-3}f_{2}^{3\omega-3}(\varepsilon)}{\sqrt{\frac{C}{6}f_{2}^{6\omega-6}
(\varepsilon)a^{6\omega-4}+\frac{P_0}{6} f_{2}^{3-3\omega}(\frac{E}{E_{Pl}})a^{3\omega-1}}}~.
\end{equation}
In the following, we shall consider the case of a power law coupling
function in the form $F(\phi)=\lambda \phi^n$. With this choice for
the function $F(\phi)$, equation (\ref{e3-27}) can be rewritten as
\begin{equation}\label{e3-28}
\frac{d\phi}{da}\phi^{n/2}=
\frac{\sqrt{\frac{C}{\lambda}}a^{3\omega-3}+\sqrt{\frac{9C}{\lambda}}\eta(\omega-1)a^{\frac{9\omega-15}{4}}}
{\sqrt{\frac{C}{6}a^{6\omega-4}+\frac{P_0}{6}a^{3\omega-1}+ C\eta(1-\omega)a^{\frac{21\omega-19}{4}}+\frac
{P_0\eta}{2}(\omega-1)a^{\frac{9\omega-7}{4}}}}~.
\end{equation}
Again, due to the lack of an analytical solution, we have restricted
ourself to a numerical solution which is plotted in figure 1 (right).
As is clear from this figure the scalar field blows up when
$a\rightarrow 0$ and tends to a constant value when $a\rightarrow
\infty$.
\begin{figure}
\begin{tabular}{c}\hspace{-1cm}\epsfig{figure=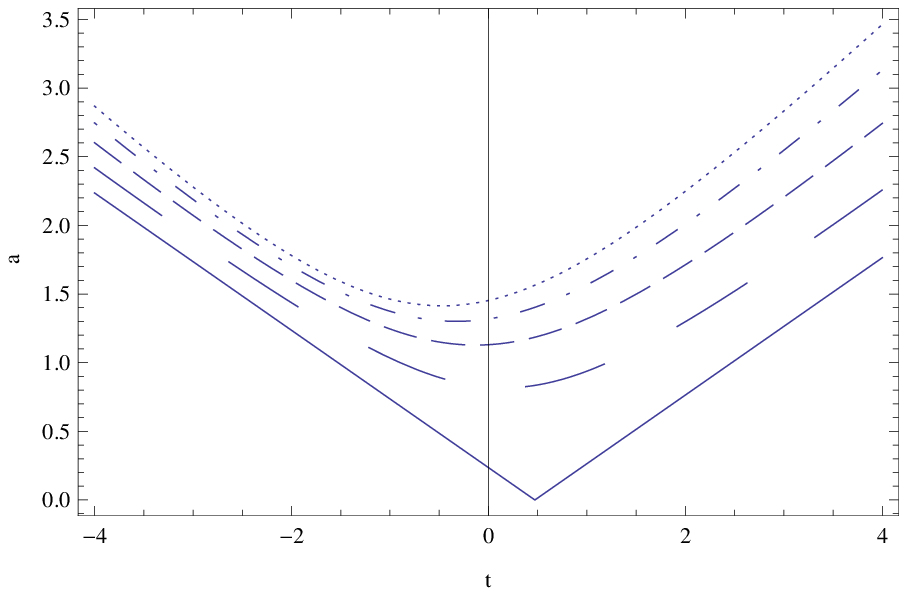,width=7cm}
\hspace{1cm} \epsfig{figure=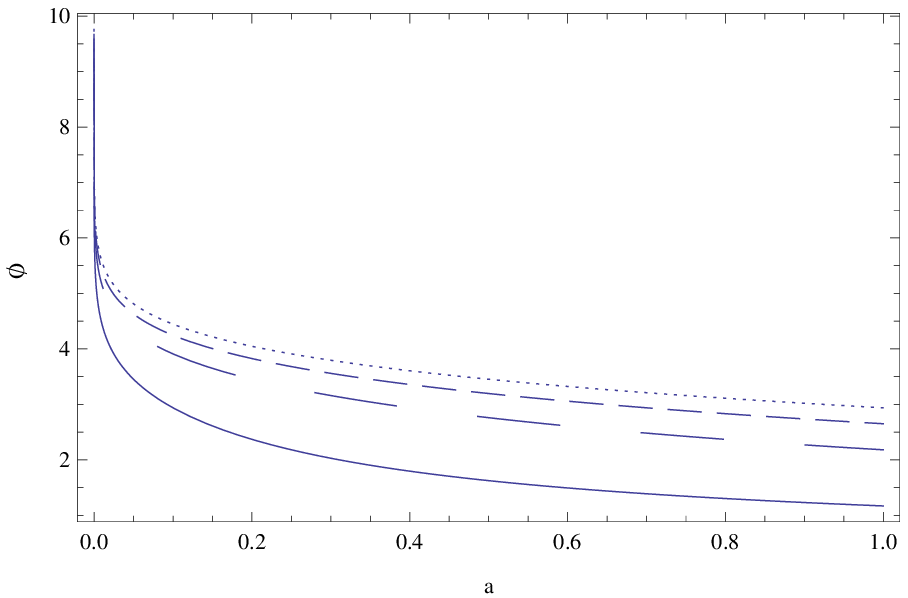,width=7cm}
\end{tabular}
\caption{\footnotesize Left: The figure shows the classical scale
factor (\ref{e3-29}) versus cosmic time for some values of $\varepsilon$:
$\varepsilon=0.2$, $\varepsilon=0.4$, $\varepsilon=0.6$,
$\varepsilon=0.8$ and $\varepsilon=1$ from down to up. Right: The
classical trajectory (\ref{e3-30}) in the plane $\phi-a$ for the
some values of $\varepsilon$: $\varepsilon=0.4$, $\varepsilon=0.6$,
$\varepsilon=0.8$ and $\varepsilon=1$ from down to up. We set numerical
values $P_0= 1$, $C=2$, $\lambda=1$, $n=2$ with the EOS parameters
$\omega=\frac{1}{3}$ for the period of the radiation dominate.}
\label{fig2}
\end{figure}
Now, let us consider an alternative case in which the average energy
of prob photons remain constant. This assumption leads to the
following solutions to the equations (\ref{e3-23}) and (\ref{e3-27})
\begin{equation}\label{e3-29}
a=\sqrt{\frac{P_{0}}{6}(1+2\varepsilon)(t-\delta)^{2}-\frac{C}{P_{0}}~\left(\frac{1-4\varepsilon}
{1+2\varepsilon}\right)},~~~~~\delta=\frac{\sqrt{6C}}{P_{0}}~\left(\frac{1-2\varepsilon}{1+2\varepsilon}
\right)~,
\end{equation}
and
\begin{equation}\label{e3-30}
\phi=\left\{
\sqrt{\frac{3}{2\lambda}}(n+2)\ln\left(\frac{\sqrt{\frac{C}{P_0}}(1-3\varepsilon)+\sqrt{\frac{C}
{P_0}(1-3\varepsilon)^{2}+a^{2}}}{a}\right)\right\}^{\frac{2}{n+2}}~,
\end{equation}where their behavior are plotted in figure 2.
Note that for the low-energy test particles i.e. $E\ll E_{Pl}$, we
have $f_{2}(\varepsilon)\rightarrow1$ and $\eta\rightarrow0$ and so
from  equations (\ref{e3-21}) and (\ref{e3-26}) we recover their
standard counterparts as
\begin{equation}\label{e3-31}
\dot{\phi}^2 F(\phi)=C a^{6\omega-6}~,
\end{equation}
and
\begin{equation}\label{e3-32}
\frac{da}{dt}=\sqrt{\frac{C}{6}
a^{6\omega-4}+\frac{P_0}{6}a^{3\omega-1} }~,
\end{equation}which with the $\omega=\frac{1}{3}$ their solutions
read as
\begin{equation}\label{e3-33}
a(t)=\sqrt{\frac{P_0}{6}(t-\delta)^2-\frac{C}{P_0}},\hspace{5mm}\delta
=\frac{\sqrt{6C}}{P_{0}}~,
\end{equation}
and
\begin{equation}\label{e3-35}
\phi(a)=\left\{\sqrt{\frac{3}{2\lambda}}(n+2)\ln\left(\frac{\sqrt{\frac{C}{P_0}}
+\sqrt{\frac{C}{P_0}+a^{2}}}{a}\right)\right\}^{\frac{2}{n+2}}~.
\end{equation}
These solutions also exhibit similar singularities as we have
already discussed when we were dealing with more general cases in
figures 1 and 2. In the next sections we will see how this picture
may be modified if one takes into account the quantum mechanical
considerations in the problem at hand.
\section{Quantization of the model}
Let us now focus attention on the quantization of the model
described above. The starting point will be the Wheeler-DeWitt (WDW)
equation extracted from the super Hamiltonian (\ref{e3-16}). Since
the lapse function plays the role of a Lagrange multiplier, so we
have the classical constraint equation ${\cal H}=0$. In order to
quantization of the minisuperspace we use the operator version of
this constraint applying on the wave function $\Psi(a,\phi,T)$. For
a flat FRW model this procedure yields
\begin{equation}\label{e4-1}
{\cal
H}\Psi(a,\phi,T)=\left[-\frac{1}{24}\frac{P_a^2}{a}+\frac{1}{4F(\phi)a^3}
P_{\phi}^2+f_2^{3\omega-3}(\varepsilon)\frac{P_T}{a^{3\omega}}\right]\Psi
(a,\phi,T)=0~.
\end{equation}
A look at this relation shows that the ordering $a^{-1}P_a^2=P_a
a^{-1}P_a$ and $F(\phi)^{-1}P_{\phi}^2=P_{\phi}F(\phi)^{-1}P_{\phi}$
makes the Hamiltonian Hermitian. So, by using of the standard
representation $(P_{a}, P_{\phi},P_{T})\rightarrow -i(\frac
{\partial}{\partial a}, \frac{\partial}{\partial \phi},
\frac{\partial}{\partial T})$, we arrive at the
Schr\"{o}dinger-Wheeler-deWitt (SWD) equation as
\begin{eqnarray}\label{e4-2}
\frac{1}{a}\frac{\partial^2\Psi(a,\phi,T)}{\partial a^2}-\frac{1}{a^{2}}
\frac{\partial\Psi(a,\phi,T)}{\partial a}-\frac{6}{a^{3}F(\phi)}\frac{\partial^2
\Psi(a,\phi,T)}{\partial \phi^2}+\nonumber\\
\frac{6}{a^{3}}\frac{F'(\phi)}{F(\phi)^2}\frac{\partial\Psi(a,\phi,T)}
{\partial\phi}-\frac{24i}{a^{3\omega}}f_2^{3\omega-3}(\varepsilon)\frac{\partial
\Psi(a,\phi,T)}{\partial T}=0~.
\end{eqnarray}
We separate the variables in this equation as
\begin{equation}\label{e4-3}
\Psi (a,\phi,T) = e^{iET} \chi (a,\phi)~,
\end{equation} by which, equation (\ref{e4-2}) becomes
\begin{eqnarray}\label{e4-4}
\frac{\partial^2\chi (a,\phi)}{\partial
a^2}-\frac{1}{a}\frac{\partial\chi (a,\phi)}{\partial a}
-\frac{6}{a^{2}F(\phi)}\frac{\partial^2\chi (a,\phi)}{\partial
\phi^2}+\nonumber\\ \frac{6}{a^{2}}
\frac{F'(\phi)}{F(\phi)^2}\frac{\partial\chi (a,\phi)}{\partial
\phi}-24Ea^{3-3\omega}f_2^{3\omega-3} (\varepsilon)\chi (a,\phi)=0~.
\end{eqnarray}
Note that here $E$ is a separation constant which due to the
assumption of constant average energy for the prob particles can be
interpreted as the scale at which the minisuperspace is searching.
The solutions of the above differential equation are separable and
may be written in the form $\psi(a,\phi)=U(a)W(\phi)$ which yields
\begin{eqnarray}\label{e4-5}
a^2\frac{d^2U(a)}{da^2}-a\frac{dU(a)}{da}+\bigg(24Ef_2^{3\omega-3}
(\varepsilon)a^{3-3\omega}+m^2\bigg)U(a)=0~,
\end{eqnarray}
and
\begin{eqnarray}\label{e4-6}
\frac{6}{F(\phi)}\frac{d^2W(\phi)}{d\phi^2}-\frac{6F'(\phi)}{F(\phi)^2}
\frac{dW(\phi)}{d\phi}+m^2W(\phi)=0~,
\end{eqnarray}
where $m$ is another constant of separation. It is seen that the
scalar field part of the wave function is not affected by the
rainbow function. On the other hand, by choosing the rainbow
function $f_{2}(\varepsilon)=(1-\frac{E}{E_{Pl}})^{-1}$, the
differential equation (\ref{e4-5}) can be rewritten as
\begin{eqnarray}\label{e4-7}
a^2\frac{d^2U(a)}{da^2}-a\frac{dU(a)}{da}+\bigg[\bigg(24E-\frac{48E^{2}}
{E_{Pl}}\bigg)a^{2}+m^2\bigg]U(a)=0,
\end{eqnarray}
whose general solution may be given in terms of the Bessel functions
as
\begin{eqnarray}\label{e4-8}
U(a)=a\left[c_1J_{\sqrt{1-m^{2}}}\left(\sqrt{24E-\frac{48E^{2}}{E_{Pl}}}a\right)+
c_2Y_{\sqrt{1-m^{2}}}\left(\sqrt{24E+\frac{48E^{2}}{E_{Pl}}}a\right)\right].
\end{eqnarray}
Also, with $F(\phi)=\lambda\phi^{n}$, equation (\ref{e4-6}) has
exact solution as
\begin{eqnarray}\label{e4-9}
W(\phi)=\phi^{\frac{n+1}{2}}\left[c_{3}J_{\frac{n+1}{n+2}}\left(\frac{m\sqrt
{6\lambda}}{3(n+2)}\phi^{\frac{n+2}{2}}\right)+c_{4}Y_{\frac{n+1}{n+2}}\left
(\frac{m\sqrt{6\lambda}}{3(n+2)}\phi^{\frac{n+2}{2}}\right)\right]~.
\end{eqnarray}
Since the Bessel function $Y_{\nu}(z)$ has not a well-defined
behavior near $z\approx0$, we may set $c_2=c_{4}=0$. Therefore, the
eigenfunctions of the SWD equation will be
\begin{equation}\label{e4-10}
\Psi_{m E}(a,\phi,T)=e^{iET}a \phi^{\frac{n+1}{2}} J_{\sqrt{1-m^{2}}}\left
(\sqrt{24E-\frac{48E^{2}}{E_{Pl}}}a\right)J_{\frac{n+1}{n+2}}\left(\frac
{m\sqrt{6\lambda}}{3(n+2)}\phi^{\frac{n+2}{2}}\right)~.
\end{equation}
We note that if we restricted ourselves to the real values of the
Bessel function's argument, i.e. $24E-\frac{48E^{2}}{E_{Pl}}\geq0$,
an energy cut-off appears as $E\leq\frac{E_{Pl}}{2}$. Now, the
general solution to the SWD equation may be written as a
superposition of its eigenfunctions, that is
\begin{equation}\label{e4-11}
\Psi(x,y,T)=\int_{E=0}^{\infty}\int_{m=0}^1 B(E)C(m)\Psi_{m
E}(x,y,T) dE dm~,
\end{equation}
where $B(E)$ and $C(m)$ are suitable weight functions to construct
the wave packets. To consider the effects of the energy's cut-off on
the above wave function, we re-scale the energy as
$dE\rightarrow(1-\frac{E}{E_{Pl}})^{-1}dE$, so that the relation
(\ref{e4-11}) takes the form
\begin{equation}\label{e4-12}
\Psi(x,y,T)=\int_{m=0}^1 (I_{1}+I_{2})C(m) dm~,
\end{equation}
in which
\begin{equation}\label{e4-13}
I_{1}= \int_{E=0}^\infty B(E)\Psi_{m E}(x,y,T) dE,~~~~~~ I_{2}=
\frac{1}{E_{Pl}}\int_{E=0}^\infty EB(E)\Psi_{m E}(x,y,T) dE~.
\end{equation}
The above integrals may be evaluated to find analytical expression
if we choose the function $B(E)$ to be a quasi-Gaussian weight
factor
\begin{equation}\label{e4-14}
B(E)=12\left(\sqrt{24E(1-\frac{2}
{E_{Pl}})}\right)^{\sqrt{1-m^2}}\exp \left(-24E(1-\frac{2}
{E_{Pl}})\right)~,
\end{equation}in which we also have used the approximation
$24E-\frac{48E^{2}}{E_{Pl}}\approx24E(1-\frac{2}{E_{Pl}})$.
By using the equalities \cite{Abr}
\begin{equation}\label{e4-17}
\int_0^\infty e^{-a
z^2}z^{\nu+1}J_{\nu}(bz)dz=\frac{b^{\nu}}{(2a)^{\nu+1}}e^{-\frac
{b^2}{4a}}~,~~~~\int_0^\infty e^{-az^2}z^{\nu+3}J_{\nu}(bz)dz=\frac
{b^{\nu}\left[-b^{2}+4a(1+\nu)\right]}{(2a)^{\nu+3}}e^{-\frac{b^2}{4a}}~,
\end{equation}
the result is
\begin{eqnarray}\label{e4-15}
I_{1}&=&(1+\frac{2}{E_{Pl}})a\phi^{\frac{n+1}{2}}J_{\frac{n+1}{n+2}}
\left(\frac{m\sqrt{6\lambda}}{3(n+2)}\phi^{\frac{n+2}{2}}\right)\left
[a^{\sqrt{1-m^{2}}}\left(2-\frac{iT}{12}-\frac{iT}{6E_{Pl}}\right)^{-1
-\sqrt{1-m^{2}}}\right. \nonumber\\ & &\left.\exp\left(-\frac{a^{2}}
{\left(4-\frac{iT}{6}-\frac{iT}{3E_{Pl}}\right)}\right)\right]~,
\end{eqnarray}
and
\begin{eqnarray}\label{e4-16}
I_{2}&=&\frac{1}{24E_{Pl}}a\phi^{\frac{n+1}{2}}J_{\frac{n+1}{n+2}}
\left(\frac{m\sqrt{6\lambda}}{3(n+2)}\phi^{\frac{n+2}{2}}\right)\left
[a^{\sqrt{1-m^{2}}}\left(2-\frac{iT}{12}-\frac{iT}{6E_{Pl}}\right)^{-3
-\sqrt{1-m^{2}}} \right. \nonumber\\ & &\left.\exp\left(-\frac{a^{2}}
{\left(4-\frac{iT}{6}-\frac{iT}{3E_{Pl}}\right)}\right)\left(a^{2}+(1+
\sqrt{1-m^{2}})\left(4-\frac{iT}{6}-\frac{iT}{3E_{Pl}}\right)\right)
\right]~.
\end{eqnarray}
By means of these relations in (\ref{e4-12}), we have
\begin{eqnarray}\label{e4-18}
\Psi(a,\phi,T)&=&\left\{(1+\frac{2}{E_{Pl}})\left(2-\frac{iT}{12}-\frac{iT}
{6E_{Pl}}\right)^{-1-\gamma}+\frac{1}{24E_{Pl}}\left(2-\frac{iT}{12}-\frac
{iT}{6E_{Pl}}\right)^{-3-\gamma} \right.\nonumber\\ & &\left .\left(-a^{2}
+(1+\gamma)\left(4-\frac{iT}{6}-\frac{iT}{3E_{Pl}}\right)\right)\right\}a^
{1+\gamma}\exp \left(-\frac{a^{2}}{\left(4-\frac{iT}{6}-\frac{iT}{3E_{Pl}}
\right)}\right)\nonumber\\ && \times \phi^{\frac{n+2}{2}}\int_{0}^{1}C(m)
J_{\frac{n+1}{n+2}}\left(\frac{m\sqrt{6\lambda}}{3(n+2)}\right) dm~.
\end{eqnarray}
To obtain an analytical closed expression for the wave function,
what remains is evaluation of the integral over $m$. At this step we
assume that the above superposition is taken over such values of $m$
for which one can use the approximation $\sqrt{1-m^{2}}=\gamma$.
Now, by choosing the weight function
\begin{equation}\label{e4-19}
C(m)=m^{\frac{2n+3}{n+2}}~,
\end{equation}where $n$ is an arbitrary constant, and using the equality
\cite{Abr}
\begin{equation}\label{e4-21}
\int_0^1 m^{r+1} J_r(zm)dm=\frac{J_{r+1}(z)}{z}~,
\end{equation}
we are led to the following expression for the (radiation dominated)
wave function
\begin{eqnarray}\label{e4-20}
\Psi(a,\phi,T)&=&2\left(\frac{\sqrt{6\lambda}}{3(n+1)}\right)^{-1}\phi^
{-\frac{1}{2}}a^{1+\gamma}\exp \left(-\frac{a^{2}}{\left(4-\frac{iT}{6}
-\frac{iT}{3E_{Pl}}\right)}\right)\nonumber\\ && \times J_{\frac{2n+3}
{n+2}}\left(\frac{m\sqrt{6\lambda}}{3(n+2)}\phi^{\frac{n+2}{2}}\right)
\left\{(1+\frac{2}{E_{Pl}})\left(2-\frac{iT}{12}-\frac{iT}{6E_{Pl}}\right)
^{-1-\gamma}\right. \nonumber\\&&\left.-\frac{a^{2}}{24E_{Pl}}\left(2-\frac{iT}
{12}-\frac{iT}{6E_{Pl}}\right)^{-3-\gamma} \right\}~.
\end{eqnarray}

\begin{figure}
\begin{tabular}{c}\hspace{-1cm}\epsfig{figure=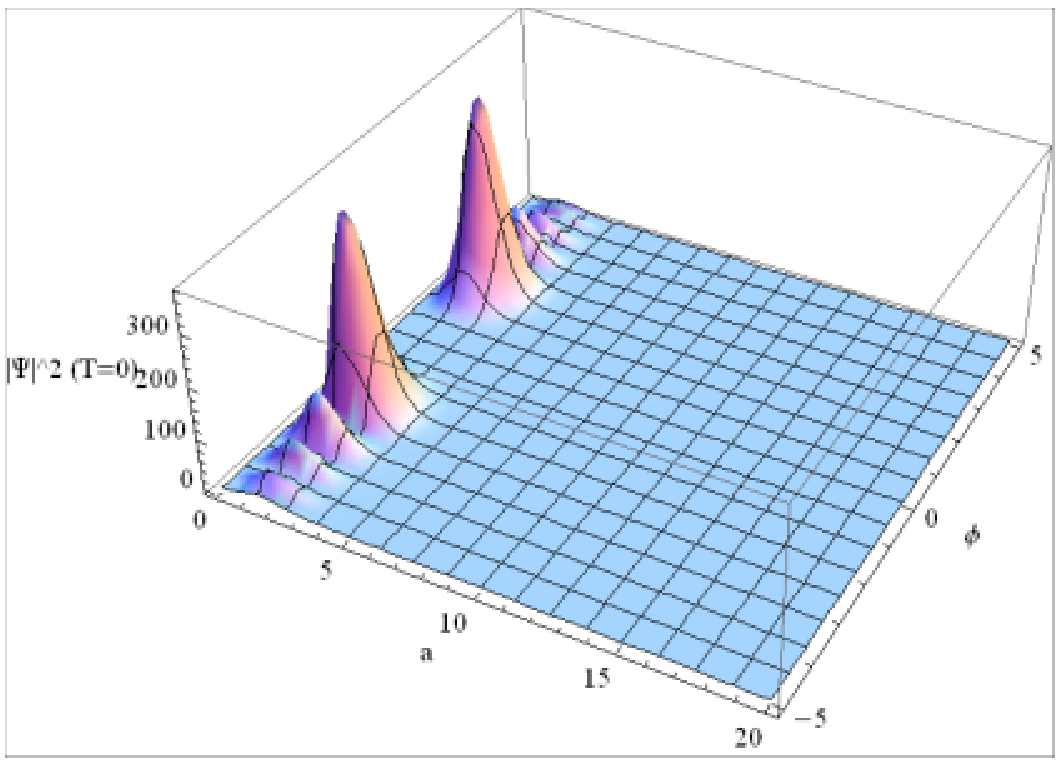,width=7cm}
\hspace{1cm} \epsfig{figure=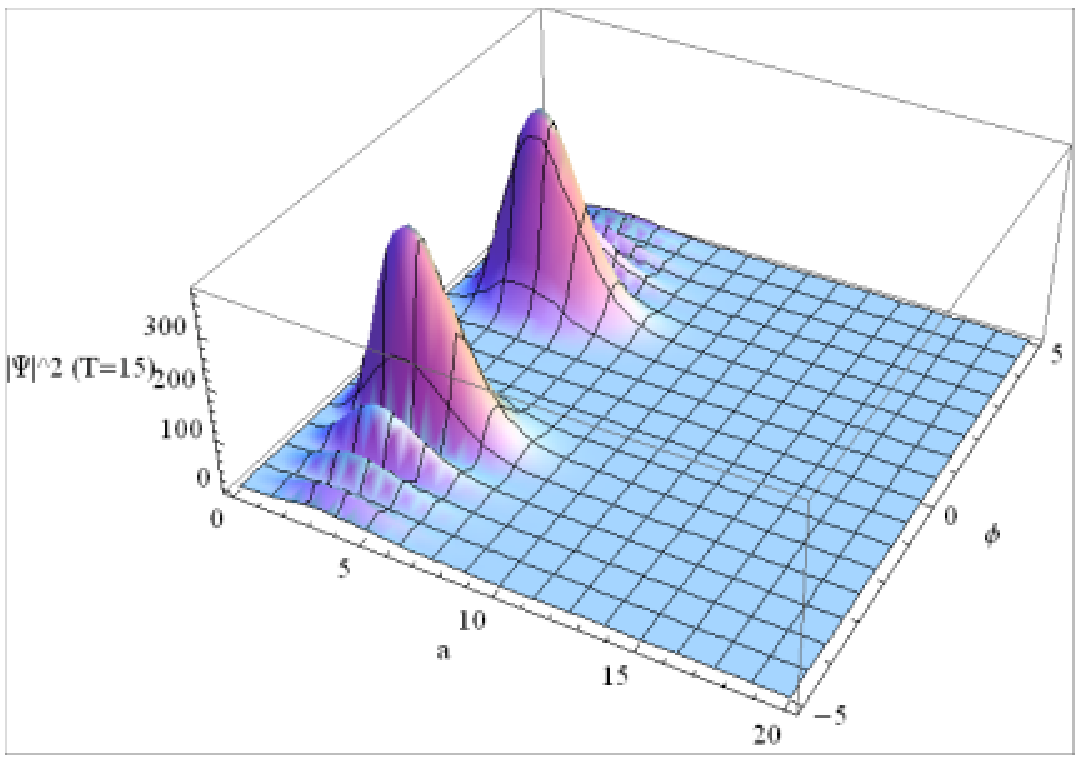,width=7cm}
\end{tabular}
\begin{tabular}{c}\hspace{-1cm}\epsfig{figure=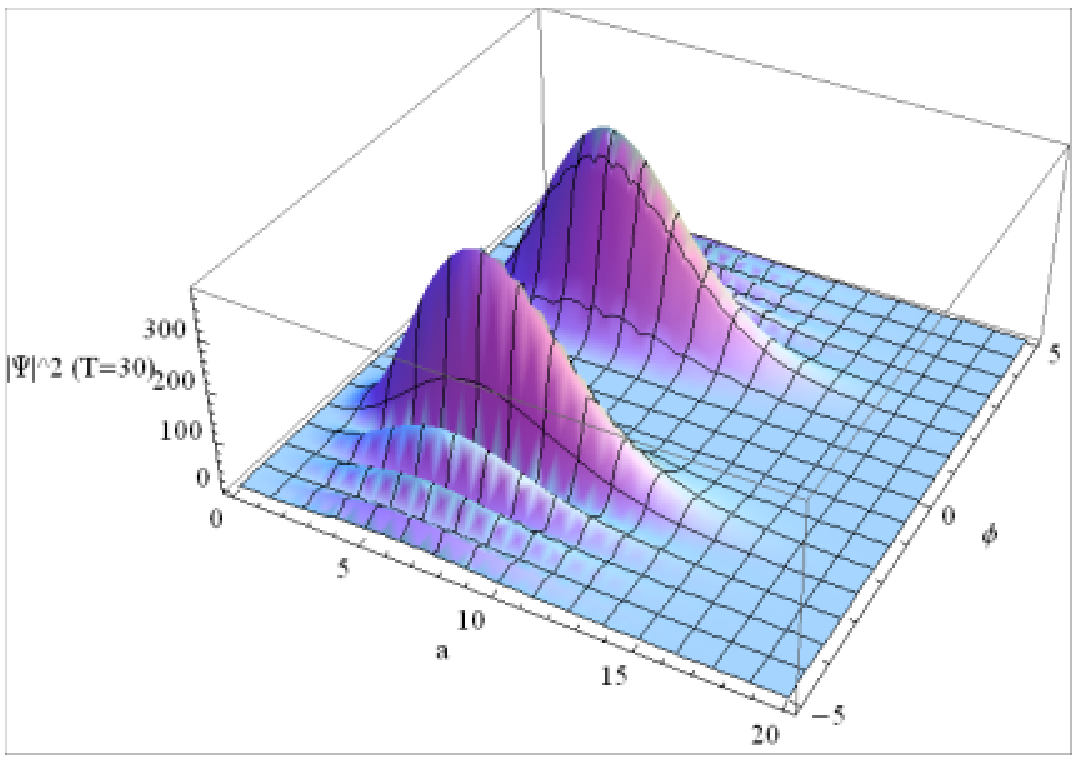,width=7cm}
\hspace{1cm} \epsfig{figure=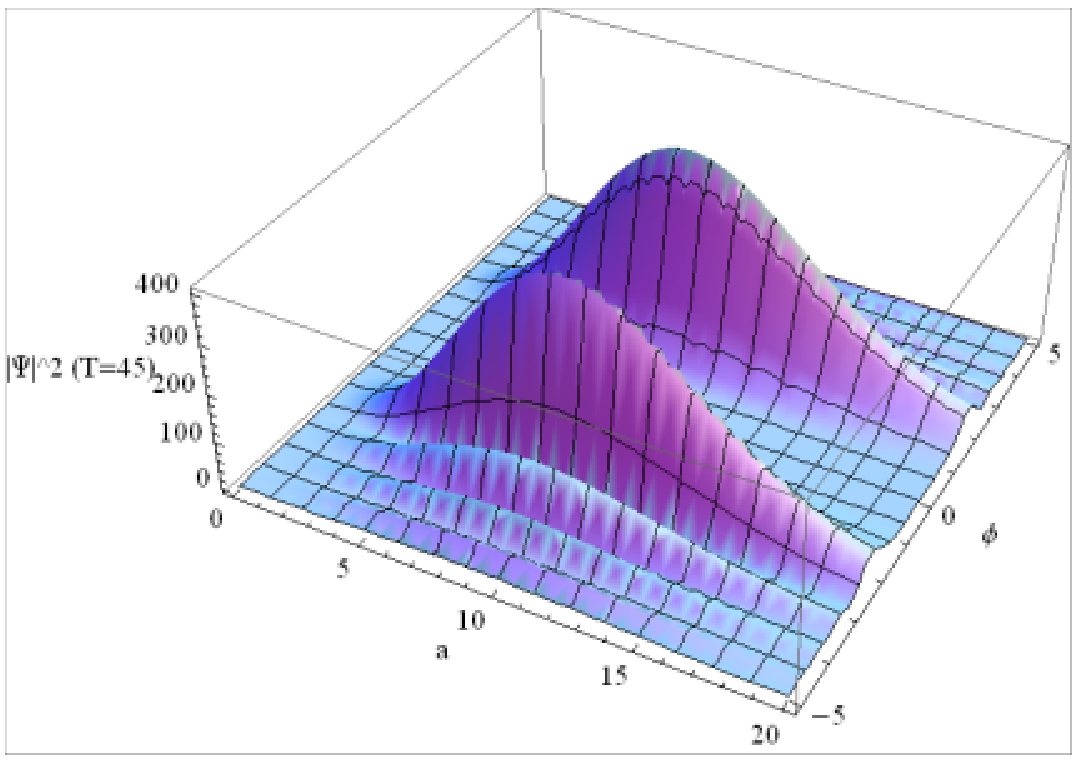,width=7cm}
\end{tabular}
\caption{\footnotesize  The probability density functions
$|\Psi(a,\phi, T)|^2$ for the wave function (\ref{e4-20}) in four
different time parameter $T=0,15,30,45$ with the numerical values
$\gamma = 0.25$, $n = 2$, $\lambda=2$ and $E_ {Pl}=1$.} \label{fig3}
\end{figure}

In figure 3, we have plotted the probability densities correspond to
the wave function (\ref{e4-20}) for typical numerical values of the
parameters. As this figure shows, at $T=0$, the wave function has
two dominant peaks in the vicinity of some non-zero values of scale
factor and scalar field. This means that the wave function predicts
the emergence of the universe from a quantum state corresponding to
one of its dominant peaks. However, the emergence of several peaks
in the wave packet may be interpreted as a representation of
different quantum states that may communicate with each other
through tunneling. In the other words, there are different possible
states from which our present universe could have evolved and
tunneled in the past, from one universe (state) to another. As time
grows, the wave packet begins to propagate in the $a$-direction, its
width becoming wider and its peaks moving with a group velocity
towards the greater values of scale factor while the values of
scalar field remain almost constant. The wave packet disperses as
time passes, the minimum width being attained at $T = 0$. As in the
case of the free particle in quantum mechanics, the more localized
the initial state at $T = 0$, the more rapidly the wave packet
disperses. Therefore, the quantum effects make themselves important
only for small enough $T$ corresponding to small $a$, as expected
and the wave function predicts that the universe will undergo into
the states with larger $a$ and an almost constant $\phi$ in its late
time evolution. If we turn off the rainbow's effects in the wave
function (see figure 1 of \cite{Babak}), after examining the same
numerical process, we verify that while the general behavior of the
wave function is repeated, the rainbow model will go into its
classical regime faster than the ordinary FRW model. This may be a
consequence of the existence of a natural cutoff as $E_{Pl}$, in the
metric.

Up to now, from our starting point to separate the variables in
(\ref{e4-3}), we assumed that the parameter $E$ is a constant
interpreted as the constat average energy of the ensemble of
photons. In other words, here the source of $E$ comes from perfect
fluid composed of photons with constant average energy and thus one
can interpret it as the scale at which the minisuperspace is
measuring. Now, we may relax this assumption and let such an energy
scale can be changed. If so, we are not allowed to identity the
separation constant (the oscillatory exponential term in equation
(\ref{e4-3})) with the energy scale by which the minisuperspace is
probing. To handle this issue, let us do a slight change in notation
and show the separation constant appeared in (\ref{e4-3}) with
$\tilde{E}$ which from now on should be distinguished from the
energy of the probing photons. So, by putting the rainbow function
into (\ref{e4-4}), equation (\ref{e4-5}) can be rewritten as
\begin{eqnarray}\label{e4-22}
a^2\frac{d^2U(a)}{da^2}-a\frac{dU(a)}{da}+\bigg(24\tilde{E}a^{2}+48\eta
\tilde{E}a+m^2\bigg)U(a)=0~.
\end{eqnarray}
Note that, only the scale factor part of $\chi(a,\phi)$ is affected
by the above assumption. The general solution of the differential
equation (\ref{e4-22}) may be written as

\begin{eqnarray}\label{e4-28}
U(a)=a^{\delta}\exp\left(\sqrt{-96\tilde{E}}a\right){\cal
J}\left(-\frac{1}{2}+\delta-\eta\sqrt{-6\tilde{E}},~
2\delta-1;~\frac{a}{2\sqrt{-96\tilde{E}}}\right),
\end{eqnarray}where $\delta=1\pm \sqrt{1-m^2}$ and ${\cal
J}(\alpha,\beta;z)$ is a degenerate hypergeometric function. For
negative values of $\tilde{E}$, $\tilde{E}<0$, this expression has
not a well-defined asymptotic behavior for large values of the
variable $a$. Thus we consider $\tilde{E}>0$, for which with
assumption $\eta\ll1$, we have
\begin{eqnarray}\label{e4-29}
U(a)=a^{\delta}\exp\left(i\sqrt{96\tilde{E}}a\right){\cal
J}\left(-\frac{1}{2}+\delta,~
2\delta-1;~\frac{-ia}{2\sqrt{96\tilde{E}}}\right).
\end{eqnarray}
By using of the following representations of the Bessel function
$J_{\nu}(z)$ in terms of the degenerate hypergeometric functions
\cite{Abr}
\begin{eqnarray}\label{e4-30}
J_{\nu}(z)= \frac{1}{\Gamma(\nu+1)}(z/2)^{\nu}\exp(-iz){\cal
J}\left(\nu+\frac{1}{2},~ 2\nu+1;~2iz\right),
\end{eqnarray}
the solution (\ref{e4-29}) reads as
\begin{eqnarray}\label{e4-31}
U(a)=\Gamma(\delta)a\left(-\frac{1}{8\sqrt{96\tilde{E}}}\right)^{1-\delta}
\exp\left(i\sqrt{96\tilde{E}}a\right)
\exp\left(\frac{-ia}{4\sqrt{96\tilde{E}}}\right)J_{\delta-1}\left
(-\frac{a}{4\sqrt{96\tilde{E}}}\right)~.
\end{eqnarray}
Therefore, the eigenfunctions of the SWD equation are given by
\begin{eqnarray}\label{e4-32}
\Psi_{m \tilde{E}}(a,\phi,T)&=&e^{i\tilde{E}T}a \phi^{\frac{n+1}{2}}
\Gamma(1\pm\sqrt{1-m^{2}})\left(-\frac{1}{4\sqrt{96\tilde{E}}}\right)^
{\mp\sqrt{1-m^{2}}} \exp\left(i\sqrt{96\tilde{E}}a\right)\nonumber\\
& & \times \exp\left(\frac{-ia}{4\sqrt{96\tilde{E}}}\right)
J_{\pm\sqrt{1-m^{2}}}\left(-\frac{a}{4\sqrt{96\tilde{E}}}\right)
J_{\frac{n+1}{n+2}}\left(\frac{m\sqrt{6\lambda}}{3(n+2)}
\phi^{\frac{n+2}{2}}\right)~.
\end{eqnarray}
As before, a superposition over the parameters $\tilde{E}$ and $m$
is needed to construct the wave function, that is
\begin{equation}\label{e4-322}
\Psi(a,\phi,T)=\int_{\tilde{E}=0}^\infty
\int_{m=0}^1B(\tilde{E})C(m)\Psi_{m \tilde{E}}(a,\phi,T).
\end{equation}
To evaluate the integral over $\tilde{E}$ and get an analytic
expression we do it under approximations $T\tilde{E}\ll1$ and
$a\sqrt{96\tilde{E}}\ll1$ which result in
$e^{i\tilde{E}T}\approx1+i\tilde{E}T$ and
$e^{i\sqrt{96\tilde{E}}a}\approx1$. By such approximations we may
offer the weight function $B(\tilde{E})$ as
\begin{eqnarray}\label{e4-33}
B(z)=768a^{\mu}\exp(-\frac{a}{4})
\exp\left(-a(z+\frac{i}{2})^{2}\right)z^{6\pm2\sqrt{1-m^{2}}},~~~~z=-\frac{1}{4\sqrt
{96\tilde{E}}},
\end{eqnarray}where $\mu$ is a positive constant. Under
this condition we are led to the final form for the wave function as
\begin{eqnarray}\label{e4-35}
\Psi(a,\phi, T)&=&\Gamma(1+\gamma)\phi^{-\frac{1}{2}}
\left(\frac{\sqrt{6\lambda}}{3(n+1)}\right)^{-1}\exp(-\frac{a}{2})\left
\{\left(\frac{1+\gamma}{2}\right)
a^{\mu-1}+\left(\frac{iT}{3072}-4\right)a^{\mu}\right\} \nonumber\\
& & \times J_{\frac{2n+3}{n+2}}
\left(\frac{m\sqrt{6\lambda}}{3(n+2)}\phi^{\frac{n+2}{2}}\right),
\end{eqnarray}in which we have taken again the integral over $m$
by approximation $\sqrt{1-m^{2}}=\gamma$ and with the weight
function (\ref{e4-19}). A qualitative investigation of the wave
function (\ref{e4-35}) shows that it has the same behavior as the
figure 3, but with different (smaller) spreading rate.

Now, having the above wave functions let us see how the classical
solutions may be recovered. In quantum cosmology, one usually
constructs a coherent wave packet with good asymptotic behavior in
the minisuperspace, peaking in the vicinity of the classical
trajectory. So, to show the correlations between classical and
quantum pattern, following the many-worlds interpretation of quantum
mechanics \cite{T}, one may calculate the time dependence of the
expectation value of the scale factor as
\begin{equation}\label{e4-36}
<a>_{T}=\frac{\int_{a=0}^{\infty}\int_{\phi=-\infty}^{\infty}
\Psi^*a\Psi da d\phi}{\int_{a=0}^{\infty}\int_{\phi=-\infty}^{\infty}
\Psi^*\Psi da d\phi}~,
\end{equation}
which yields
\begin{eqnarray}\label{e4-37}
<a>_{T}=\frac{A_{1}\bigg(16+\frac{T}{36}+\frac{T}{9E_{Pl}}\bigg)-A_{2}\bigg(16+
\frac{T}{36}+\frac{T}{9E_{Pl}}\bigg)^{2}}{A_{3}\bigg(16+\frac{T}{36}+\frac{T}
{9E_{Pl}}\bigg)^{1/2}-A_{4}\bigg(16+\frac{T}{36}+\frac{T}{9E_{Pl}}\bigg)^{-5/2}}~,
\end{eqnarray}with
\begin{equation}\label{e4-38}
A_{1}=\bigg(\frac{3}{2}+\frac{3}{8E_{Pl}}\bigg),~~~A_{2}=\frac{1}{32E_{Pl}},~~~
A_{3}=\sqrt{\frac{9\pi}{2}}A_{1},~~~A_{4}=\frac{2}{17}A_{2}~,
\end{equation} for the wave function (\ref{e4-20}) with $\gamma=1$,
and
\begin{eqnarray}\label{e4-38}
<a>_{T}=\frac{(\frac{1+\gamma}{2})^{2}+\left(\frac{T^{2}}{1572864}+96\right)-
8(1+\gamma)}{(\frac{1+\gamma}{2})^{2}+\left(\frac{T^{2}}{4718592}+32\right)-
4(1+\gamma)}~,
\end{eqnarray}for the wave function (\ref{e4-35}). In view of the existence of
singularities, these expectation values never vanish, showing that
the corresponding quantum states are nonsingular. Indeed, the
expression (\ref{e4-37}) and (\ref{e4-38}) represent a bouncing
universe with no singularity where its late time behavior is almost
the same as classical solution shown in figure 1. Also, the
expectation value of the scalar field reads as
\begin{equation}\label{e4-39}
<\phi>_{T}=\frac{\int_{a=0}^{\infty}\int_{\phi=-\infty}^{\infty}\Psi^*\phi\Psi da d\phi}
{\int_{a=0}^{\infty}\int_{\phi=-\infty}^{\infty}
\Psi^*\Psi da d\phi},
\end{equation}
with the result
\begin{eqnarray}\label{e4-40}
<\phi>_{T}&=&\frac{\int_{-\infty}^{+\infty}z^{-\frac{n}{n+2}}
J_{\frac{2n+3}{n+2}}^{^{\star}}\left
(\frac{\sqrt{6\lambda}}{3(n+2)}z\right)\times
J_{\frac{2n+3}{n+2}}\left(\frac{\sqrt{6\lambda}}{3(n+2)}
z\right)dz}{\int_{-\infty}^{+\infty}z^{-1}
J_{\frac{2n+3}{n+2}}^{^{\star}}\left(\frac{\sqrt{6\lambda}}
{3(n+2)}z\right)\times
J_{\frac{2n+3}{n+2}}\left(\frac{\sqrt{6\lambda}}{3(n+2)}z\right)dz}=\nonumber\\
& &
\frac{\Gamma(\frac{n}{n+1})\Gamma(\frac{3n+5}{n+2})\Gamma(2n+4)}{2^{\frac{n}{n+2}}\left(\Gamma(\frac{n+1}
{n+2})\right)^{2}\Gamma(\frac{2n+3}{n+2})\Gamma(\frac{3n+4}{n+2})}\left(\frac{\sqrt{6\lambda}}{3(n+2)}
\right),
\end{eqnarray}
where $z=\phi^{\frac{n+2}{2}}$. We see that the expectation value of
$\phi$ is a time independent constant which is just the behavior
predicted by the wave function of the SWD equation in figure 3.

\section{Bohmian interpretation of quantum cosmological model}
In this section we are going to study the classical behavior of the
dynamical variables $a$ and $\phi$ in the framework of Bohmian
quantum mechanics, according which the general form of the wave
function may be written as
\begin{equation}\label{e5-1}
\Psi(a,\phi,T)=\Omega (a,\phi,T)e^{iS(a,\phi,T)},
\end{equation}
where $\Omega$ and $S$ are some real function. Substitution of this
expression into the SWD equation (\ref{e4-2}) leads to the
continuity equation

\begin{equation}\label{e5-2}
2a^2\frac{\partial \Omega}{\partial a}\frac{\partial S}{\partial a}+a^2\Omega\frac{\partial^2 S}{\partial a^2}
+a\Omega \frac{\partial S}{\partial a}-\frac{12}{F(\phi)}\frac{\partial \Omega}{\partial \phi}\frac{\partial S}
{\partial \phi}-\frac{6}{F(\phi)}\Omega \frac{\partial^2 S}{\partial \phi^2}+6\frac{F'(\phi)}{F(\phi)^2}\Omega
\frac{\partial S}{\partial \phi}-24a^{2}\frac{\partial \Omega}{\partial T}=0,
\end{equation}
and the modified Hamilton-Jacobi equation
\begin{equation}\label{e5-3}
-\frac{1}{24}\frac{1}{a}\left(\frac{\partial S}{\partial
a}\right)^2+\frac{1}{4F(\phi)a^3}\left(\frac{\partial S} {\partial
\phi}\right)^2+\frac{1}{a}\left(\frac{\partial S}{\partial
T}\right)+{\cal Q}=0,
\end{equation}
in which the quantum potential ${\cal Q}$ is defined as
\begin{equation}\label{e5-4}
{\cal Q}=\frac{1}{24a\Omega}\frac{\partial^2 \Omega}{\partial a^2}+\frac{1}{24 a^2 \Omega}\frac{\partial \Omega}
{\partial a}-\frac{1}{4a^3 F(\phi)\Omega}\frac{\partial^2 \Omega}{\partial \phi^2}+\frac{F'(\phi)}{4a^3F(\phi)^2
\Omega}\frac{\partial \Omega}{\partial \phi}.
\end{equation}
In order to determine of the functions $\Omega$ and $S$ let us
rewrite the wave function (\ref{e4-20}) in the form
\begin{eqnarray}\label{e5-5}
\Psi(a,\phi,T)&=&2\left(\frac{\sqrt{6\lambda}}{3(n+1)}\right)^{-1}
\phi^{-\frac{1}{2}}a^{2}\exp\left(-\frac{4a^{2}}{\bigg(16+\frac{T^{2}}{36}(1+\frac{2}{E_{Pl}})^{2}\bigg)}\right)
\exp\bigg(-\frac{\frac{iT}{6}(1+\frac{2}{E_{Pl}})a^{2}}{\bigg(16+\frac{T^{2}}{36}(1+\frac{2}{E_{Pl}})^{2}\bigg)}
\bigg) \nonumber\\ && \times J_{\frac{2n+3}{n+2}}\left(\frac{m\sqrt
{6\lambda}}{3(n+2)}\phi^{\frac{n+2}{2}}\right)
\left\{\frac{(1+\frac{2}{E_{Pl}})}{\bigg(4+\frac{T^{2}}{144}(1+\frac{2}{E_{Pl}})^{2}\bigg)}e^{2i\theta}\right.
\nonumber\\ &
&\left.+\frac{1}{12E_{Pl}}.~\frac{e^{5i\theta}}{\bigg(16+\frac{T^{2}}{36}(1+\frac{2}{E_{Pl}})^{2}
\bigg)^{2}\bigg(4+\frac{T^{2}}{144}(1+\frac{2}{E_{Pl}})^{2}\bigg)^{1/2}}\right.\nonumber\\&
&\left.-\frac{1}{24E_{Pl}}.~\frac{a^{2}e^{4i\theta}}{\bigg(4+\frac{T^{2}}{144}(1+\frac{2}{E_{Pl}})
^{2}\bigg)^{2}}\right\},
\end{eqnarray}
where $\theta=\tan^{-1}\bigg(\frac{T}{24}(1+\frac{2}{E_{Pl}})\bigg)$
and also we fixed $\gamma=1$. The above expression seems to be too
complicated that can be decomposed into the form (\ref{e5-1}).
However, for the small values of the the parameter $T$, we may use
the approximation $e^{2i\theta}\approx e^{4i\theta}\approx
e^{5i\theta}$, by means of which we obtain
\begin{eqnarray}\label{e5-6}
\Omega&=&2\left(\frac{\sqrt{6\lambda}}{3(n+1)}\right)^{-1}\phi^{-\frac{1}{2}}a^{2}\exp\left(-\frac{4a^{2}}
{\bigg(16+\frac{T^{2}}{36}(1+\frac{2}{E_{Pl}})^{2}\bigg)}\right)J_{\frac{2n+3}{n+2}}\left(\frac{m\sqrt
{6\lambda}}{3(n+2)}\phi^{\frac{n+2}{2}}\right) \nonumber\\
& & \times\left\{\frac{(1+\frac{2}{E_{Pl}})}{\bigg(4+
\frac{T^{2}}{144}(1+\frac{2}{E_{Pl}})^{2}\bigg)}\right.\left.+\frac{1}{12E_{Pl}}.~\frac{1}{\bigg(16+\frac{T^{2}}{36}
(1+\frac{2}{E_{Pl}})^{2}\bigg)^{2}
\bigg(4+\frac{T^{2}}{144}(1+\frac{2}{E_{Pl}})^{2}\bigg)^{\frac{1}{2}}}\right. \nonumber\\
& &\left.-\frac{1}{24E_{Pl}}.~\frac{a^{2}}{\bigg(4+\frac{T^{2}}{144}(1+\frac{2}{E_{Pl}})^{2}\bigg)^{2}}\right\}~,
\end{eqnarray}
and
\begin{equation}\label{e5-7}
S=-\frac{\frac{T}{6}(1+\frac{2}{E_{Pl}})a^{2}}{\bigg(16+\frac{T^{2}}{36}(1+\frac{2}{E_{Pl}})
^{2}\bigg)}+2\theta~.
\end{equation}
In this interpretation the classical trajectories, which determine the behavior of the
scale factor and scalar field are given by
\begin{equation}\label{e5-8}
P_a=\frac{\partial S}{\partial a}= -
\frac{\frac{T}{3}(1+\frac{2}{E_{Pl}})a}{\bigg(16+\frac{T^{2}}{36}(1+\frac{2}{E_{Pl}})^{2}
\bigg)},~~~~~P_{\phi}=\frac{\partial S}{\partial \phi}=0.
\end{equation}
Using the expressions for $P_a$ and $P_{\phi}$ in (\ref{e3-19}) and
the rainbow function $f_{2}=(1-\frac{E}{E_{Pl}})^{-1}$, after
integration and up to the order of   $\frac{1} {E_{Pl}^{2}}$, we
arrive at the following expressions
\begin{equation}\label{e5-9}
a(T)=a_0\left[16+(\frac{1}{36}+\frac{1}{9E_{pl}}) T^2\right]^{\left
(\frac{1}{2}+\frac{2}{E_{Pl}}+\frac{E}{E_{Pl}}\right)},
~~~~~\phi(T)=\mbox{const.},
\end{equation}
where $a_0$ is an integration constant. It is clear that these
solutions have the same behavior as the expectation values computed
in (\ref{e4-37}), (\ref{e4-38}) and (\ref{e4-40}) and like those are
free of singularity. The origin of the singularity avoidance in
Bohmian interpretation may be understood by the existence of the
quantum potential which corrects the classical behavior near the
classical singularity.

Recall that scale factor (\ref{e5-9}) is devoted to period of
radiation dominated full of the prob photons which during the
exploring the minisuperspace, their average energy remain nearly
constant. Finally, by combining the expressions (\ref{e5-9}) and
(\ref{e5-6}), we have drawn the behavior of $\Omega^{2}$ in terms of
scale factor $a$ for given values of numerical parameters.
\begin{figure}
\begin{tabular}{c}\hspace{1cm}\epsfig{figure=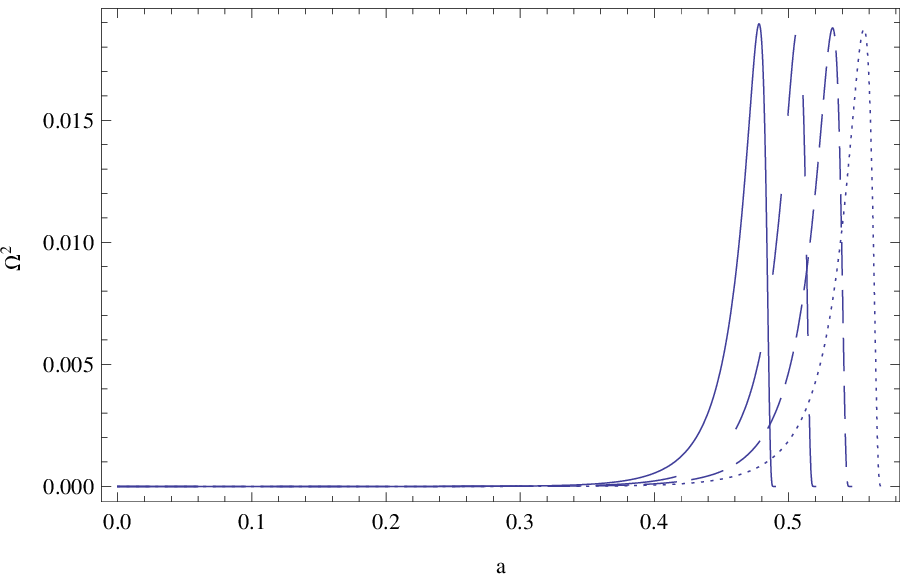,width=7cm}
\hspace{1cm}\epsfig{figure=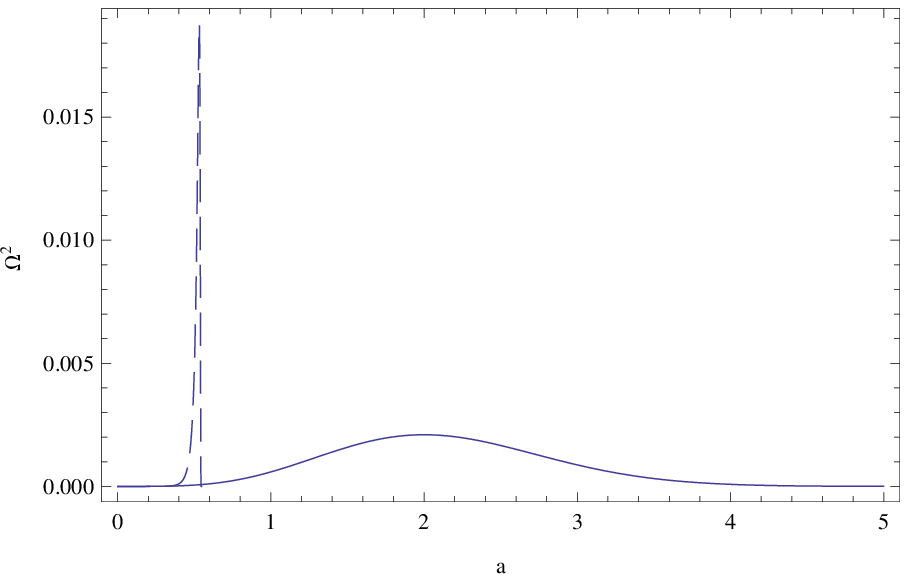,width=7cm}
\end{tabular}
\caption{\footnotesize Left: The probability density function
$\Omega^2$ in (\ref{e5-6}) versus scale factor $a$. The figure is
plotted for different values of the minisuperspace prob energy
$E=0.25$, $E=0.5$, $E=0.75$ and $E=1$ from left to right. Right: The
comparison of $\Omega^2-a$ for two difference cases: FRW metric
(solid curve) and rainbow metric (small dashed curve). We also set
the numerical values $n = 2$, $\lambda = 10$, $\phi=5$ and
$E_{Pl}=1$.} \label{fig4}
\end{figure}
Figure 4 (left) reflects the fact that in the context of Bohmian
quantum mechanics it is possible for our current universe (with
initial rainbow FRW metric) to appear from a non-zero value of scale
factor. This figure shows that as the energy of prob particles (in
particular photons) increases, the peak of $\Omega^{2}$ moves
towards greater scale factor. Noting that in limit
$E_{Pl}\rightarrow\infty$ the rainbow FRW metric is converted to
standard FRW metric, a comparison of these two cases is also plotted
in figure 4 (right) from which we see that the peak of $\Omega^2$ in
FRW background emerges in a larger scale factor in comparison with
its rainbow counterpart. However, in rainbow background the
probability of the emergence of the universe from a non-zero value
scale factor is greater and its corresponding $\Omega^2$ is sharper
and more localized.

\section{Summary and conclusions}
In this paper we have investigated the classical and quantum
dynamics of a scalar-rainbow metric cosmological model coupled to a
perfect fluid in the context of the Schutz' representation. By means
of the Schutz' formalism for perfect fluid we are able to introduce
the only remaining matter degree of freedom as a time parameter in
the model. Due to the rainbow FRW background this result has
released via a particular energy-dependent gauge fixing. In terms of
this time parameter, and in the framework of the Hamiltonian
formalism, we have obtained the corresponding classical cosmology by
evaluating the dynamical behavior of the cosmic scale factor and the
scalar field. We have seen that the classical evolution of the
universe, with the rainbow geometry background, represents a late
time expansion coming from a big-bang singularity. This is while
that if the average prob energy of minisuperspace remains almost
constant, there may be possibility of an early bounce (rather than
big-bang) from a non-zero value of scale factor. This outcome is
interesting from the cosmological standpoint in the sense that based
on a classical picture a late time expansion does not come from an
initial singularity. We then dealt with the quantization of the
model in which we saw that the classical singular behavior will be
modified.  In the quantum model, we showed that the SWD equation in
the presence of rainbow FRW metric can be separated and its
eigenfunctions can be obtained in terms of analytical functions. In
this regards, the wave functions (\ref{e4-20}) and (\ref{e4-35}) are
achieved based on  two different assumptions on minisuperspace prob
energy. Generally, both the wave functions display patterns in which
there are two possible quantum states from which our current
universe could have evolved and tunneled in the past from one state
to another. The wave function (\ref{e4-20}) has been derived with
the assumption that the average prob energy of minisuperspace
remains almost constant. Time evolution pattern of this wave
function in figure 3 demonstrated that it moves along the larger
$a$-direction whereas the scalar field $\phi$ remains constant and
does not change its shape.  As time passes, our results indicated
that the wave packets disperse and the minimum width being attained
at $T = 0$, which means that the quantum effects are important for
small enough $T$, corresponding to small $a$. The wave function
(\ref{e4-35}), on the other hand, has been obtained with this idea
that the minisuperspace prob energy does not stays constant. The
avoidance of classical singularities due to quantum effects, and the
recovery of the classical dynamics of the universe are another
important topics of our quantum presentation of the model from which
the time evolution of the expectation values of scale factor and
scalar field along with their Bohmian counterparts have been
assessed. We verified that a bouncing singularity-free universe is
obtained in both cases. Finally, by analyzing the probability
density function $\Omega^2$ versus scale factor $a$ we found that in
the rainbow geometric background the probability of the emergence of
the universe from a non-zero scale factor is greater than its FRW
metric counterpart.

\end{document}